\documentclass[journal]{IEEEtran}

\ifCLASSINFOpdf
\else
   \usepackage[dvips]{graphicx}
\fi

\usepackage{url}
\usepackage{graphicx}
\usepackage{xcolor}
\usepackage{cite}
\usepackage[colorlinks=true]{hyperref}
\usepackage{amsmath}
\usepackage{amssymb}
\usepackage{booktabs}
\usepackage{comment}
\usepackage{multirow}
\usepackage{amsbsy}
\usepackage{subcaption}

\newcommand{\tit}[1]{\smallbreak\noindent\textbf{#1.}}

\begin{document}

\title{Time-Frequency Analysis of Variable-Length\\WiFi CSI Signals for Person Re-Identification}

\author{Chen Mao, Chong Tan, \IEEEmembership{Member, IEEE}, Jingqi Hu, and Min Zheng, 
\thanks{}
\thanks{}
}

\markboth{IEEE Signal Processing Letters}
{Cartella \MakeLowercase{\textit{et al.}}: Unveiling the Truth: Exploring Human Gaze Patterns in Fake Images}
\maketitle

\begin{abstract}
Person re-identification (ReID), as a crucial technology in the field of security, plays an important role in security detection and people counting. Current security and monitoring systems largely rely on visual information, which may infringe on personal privacy and be susceptible to interference from pedestrian appearances and clothing in certain scenarios. Meanwhile, the widespread use of routers offers new possibilities for ReID. This letter introduces a method using WiFi Channel State Information (CSI), leveraging the multipath propagation characteristics of WiFi signals as a basis for distinguishing different pedestrian features. We propose a two-stream network structure capable of processing variable-length data, which analyzes the amplitude in the time domain and the phase in the frequency domain of WiFi signals, fuses time-frequency information through continuous lateral connections, and employs advanced objective functions for representation and metric learning. Tested on a dataset collected in the real world, our method achieves 93.68\% mAP and 98.13\% Rank-1.
\end{abstract}


\IEEEpeerreviewmaketitle

\section{Introduction}

\IEEEPARstart{P}{erson} ReID is a key task in the field of security and surveillance, essentially an information retrieval technology aimed at identifying the same person target across a cross-sensor perception network under various backgrounds. 
In recent years, this task has attracted widespread attention from the academic and industrial communities due to its potentially significant applications in large-scale surveillance networks, making it of great research impact and practical value.


A significant amount of ReID work\cite{wang2018learning, 2019Bags, ye2021deep, 2020FastReID, 10253470, 10341338, dong2024multi}  focuses on feature extraction and analysis from images. However, these purely visual solutions face challenges due to limitations imposed by clothing, occlusion, and color, among other factors. Considering that visual solutions are easily affected by inadequate lighting, while WiFi signals can propagate in non-line-of-sight and even through walls, and taking into account the need for privacy protection in practical applications and the widespread presence of WiFi sensing equipment in indoor environments, we turn to using WiFi CSI data as the information source for person ReID.



\begin{figure}[t!]
    \centering
    \includegraphics[width=\linewidth]{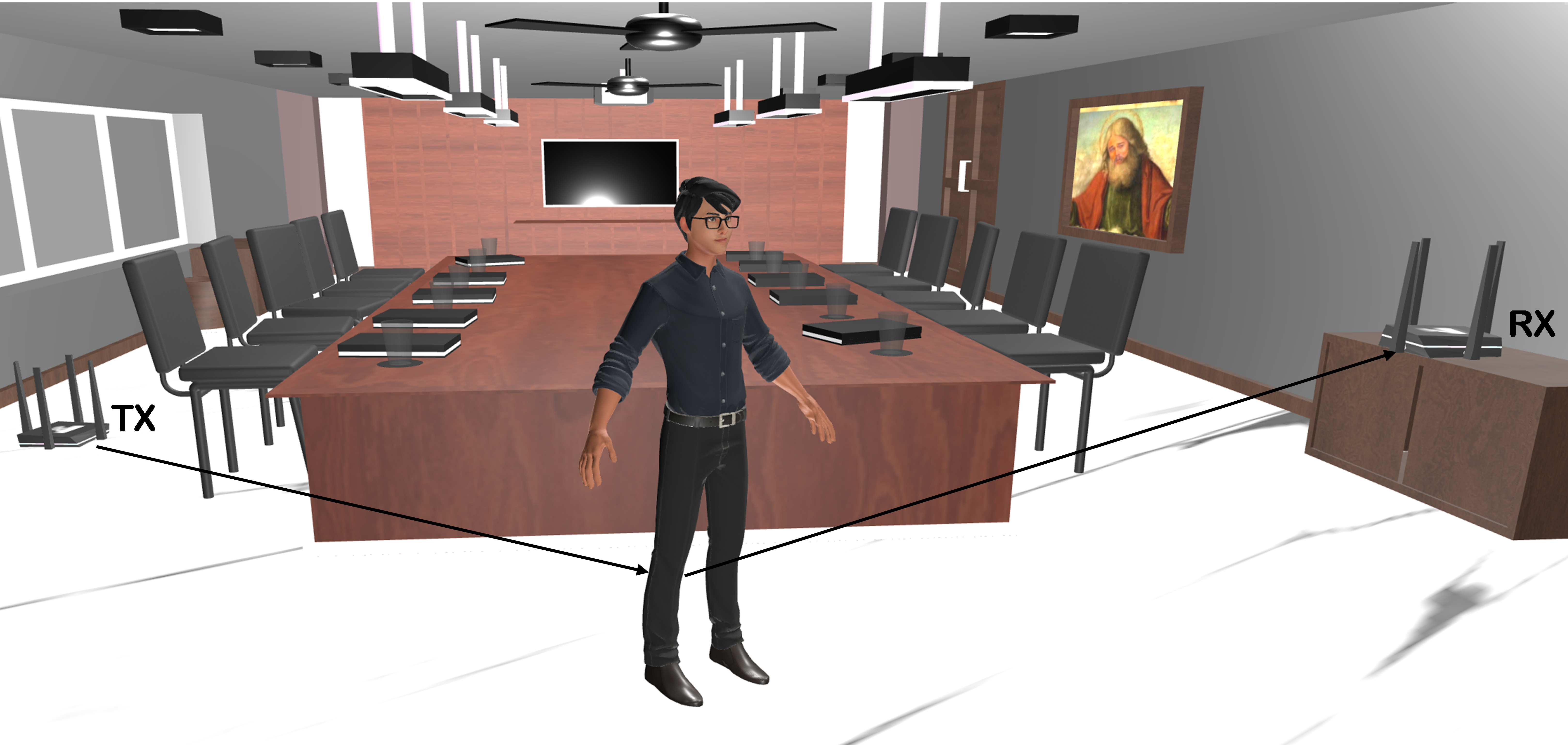}
    \caption{Schematic diagram of WiFi signal propagation to the human body.}
    \label{fig:wifi_scene}
    \vspace{-.35cm}
\end{figure}

As shown in Fig.\ref{fig:wifi_scene}, in the complex physical environment of wireless communication, when WiFi signals encounter pedestrians during propagation, the signals undergo various physical phenomena such as reflection, diffraction, and scattering on the human surface, thereby affecting the signal propagation path and its final state. These complex signal changes can be accurately captured and described through CSI, which reflects the characteristics of each subcarrier, including their amplitude attenuation and the phase shift caused by multipath effects, constituting the fine-grained features of signal propagation.

Leveraging these characteristics of WiFi CSI information, scholars begin to explore the possibilities of using WiFi CSI for advanced information processing and environmental sensing. In the area of human pose recognition\cite{li2021two, zhou2022csi, 2018Through}, minor changes in WiFi signals can be analyzed to identify human actions. For indoor positioning, WiFi CSI analysis\cite{zhang2022tips, 2015PhaseFi} provides more accurate location information than traditional positioning technologies. There is also some research on using WiFi for person re-identification and recognition\cite{2019Wi, wang2019wi, 9157423}. Some multimodal methods\cite{ren2022poster, ren2023person} combine vision and wifi signals to try to further break the bottleneck of single-modal methods.


In this letter, we propose a WiFi signal-based ReID method that makes full use of existing routers in real-world scenarios. For variable-length WiFi signal feature extraction, we introduce a concise encoding mechanism that allows the model to perceive both temporal and frequency dimensions of the signal. We employ a dual-stream network to separately process the amplitude in the time domain and the phase in the frequency domain, utilizing continuous lateral connections after each pair of encoder\cite{vaswani2017attention} outputs to fully exploit the time-frequency characteristics of WiFi signals. This approach generates more comprehensive person feature representations. We apply advanced loss functions to both the network's final predictions and the cascaded feature layers, effectively accomplishing classification-based representation learning and feature distance-based metric learning. This WiFi signal analysis-based person re-identification technology provides an innovative security application solution, enabling effective personnel monitoring and management without infringing on personal privacy.

\section{Proposed Method}


\subsection{Data Preprocessing}


Currently, most methods\cite{2019Free, zhou2022csi, wang2019wi} for analyzing information based on CSI only utilize the amplitude response of CSI. However, the calibrated CSI phase is more sensitive to changes in dynamic objects and can provide more detailed information on signal propagation path changes and multipath effects\cite{2018Mitigation, 2021Dynamic}. 

Due to the cyclical nature of the phase, and the collected signals are subject to noise interference from the surrounding environment and hardware errors during transmission. It is necessary to unwrap the CSI phase and then apply a linear transformation to eliminate phase shifts caused by carrier frequency errors and clock synchronization errors. The phase of the \(i\)-th subcarrier can be represented as:
\begin{equation}
\tilde{\phi}_i = \phi_i - \frac{2\pi k_i}{N}\delta + \beta + Z
\end{equation}

where $\phi_i$ is the original phase value information, $\delta$ is the timing offset of carrier frequency offset, $\beta$ is the unknown phase offset, and $Z$ is phase noise. $k_i$ is the index of the $i$-th subcarrier, and $N$ is the FFT size.
Assuming that the measurement noise \( Z \) is small, the errors due to the offsets \( \delta \) and \( \beta \) can be eliminated through a linear transformation:
\begin{equation}
a = \frac{\tilde{\phi}_i - \tilde{\phi}_1}{k_n - k_1} = \frac{\phi_i - \phi_1}{k_n - k_1} - \frac{2\pi\delta}{N}
\end{equation}
\begin{equation}
b = \frac{1}{n} \sum_{1 \leq j \leq n} \tilde{\phi}_j = \frac{1}{n} \sum_{1 \leq j \leq n} \phi_j - \frac{2\pi\delta}{nN} \sum_{1 \leq j \leq n} k_j + \beta
\end{equation}


where \(a\) represents the slope obtained from the linear fitting algorithm, and \(b\) is the intercept. Assuming that the subcarrier indices are symmetric such that \(\sum_{j=1}^{n} k_j = 0\), the errors due to \(\delta\) and \( \beta \) can be eliminated by using \(\phi_i - a k_i - b\). Then we obtain the calibrated phase, which differs from the true phase by a constant multiple \(c_i\) related to the frequency. 
\begin{equation}
\sigma^2_{\tilde{\phi}_i} = c_i \sigma^2_{\phi_i}, \quad c_i = 1 + 2 \frac{k_i^2}{(k_n - k_1)^2} + \frac{1}{n}
\end{equation}

\begin{figure}[t!]
    \centering
    \includegraphics[width=\linewidth]{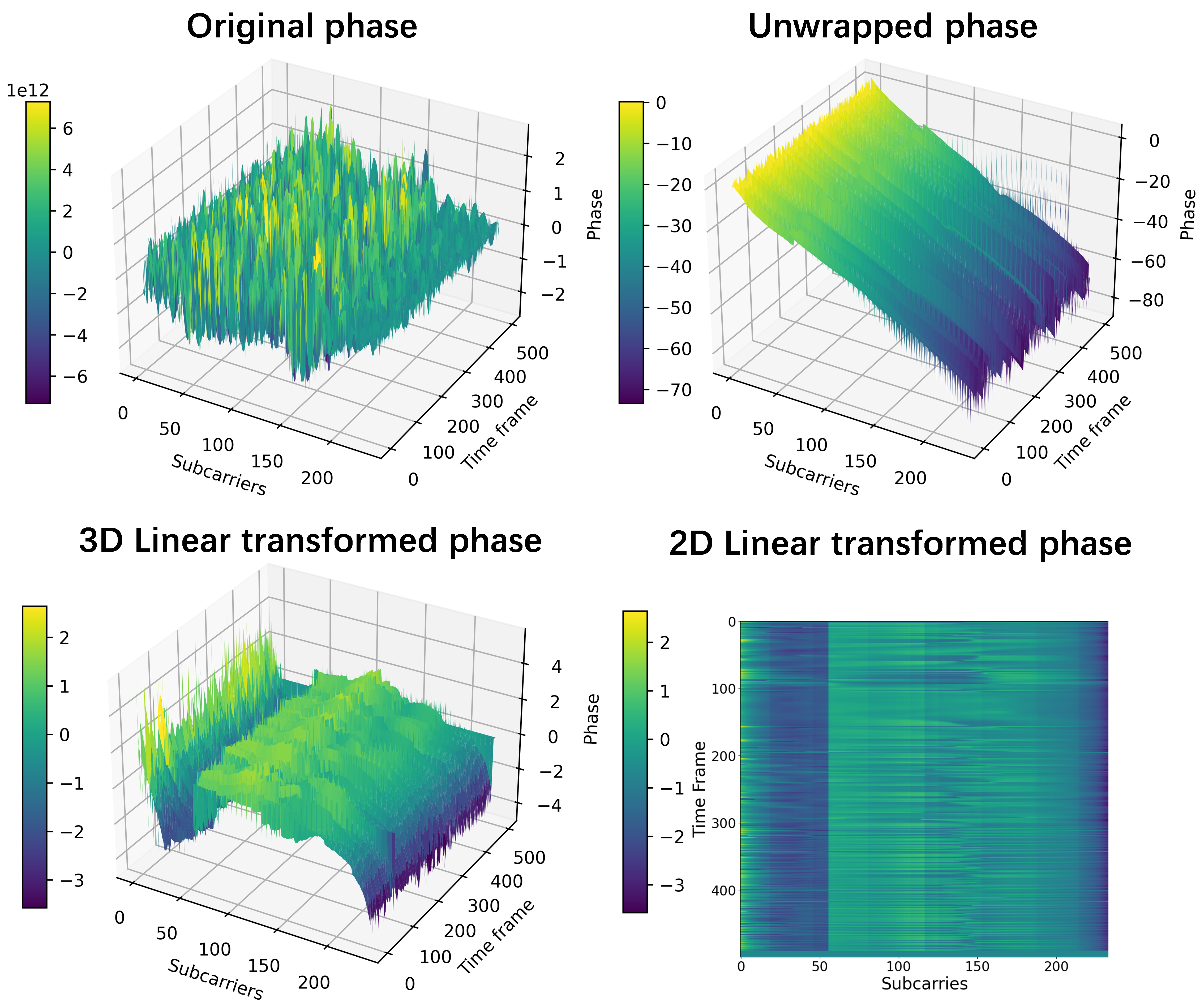}
    \caption{Original, unwrapped and linear transformed phase visualization.}
    \label{fig:phase}
    \vspace{-.35cm}
\end{figure}

The visualization of the original, the unwrapped, and the linear transformed phase are shown in Fig. \ref{fig:phase}. The phase before processing appears to be without any discernible pattern, while the processed phase is concentrated and distributed more uniformly, demonstrating the effectiveness of phase correction. 

\begin{figure*}[t!]
    \centering
    \includegraphics[width=\linewidth]{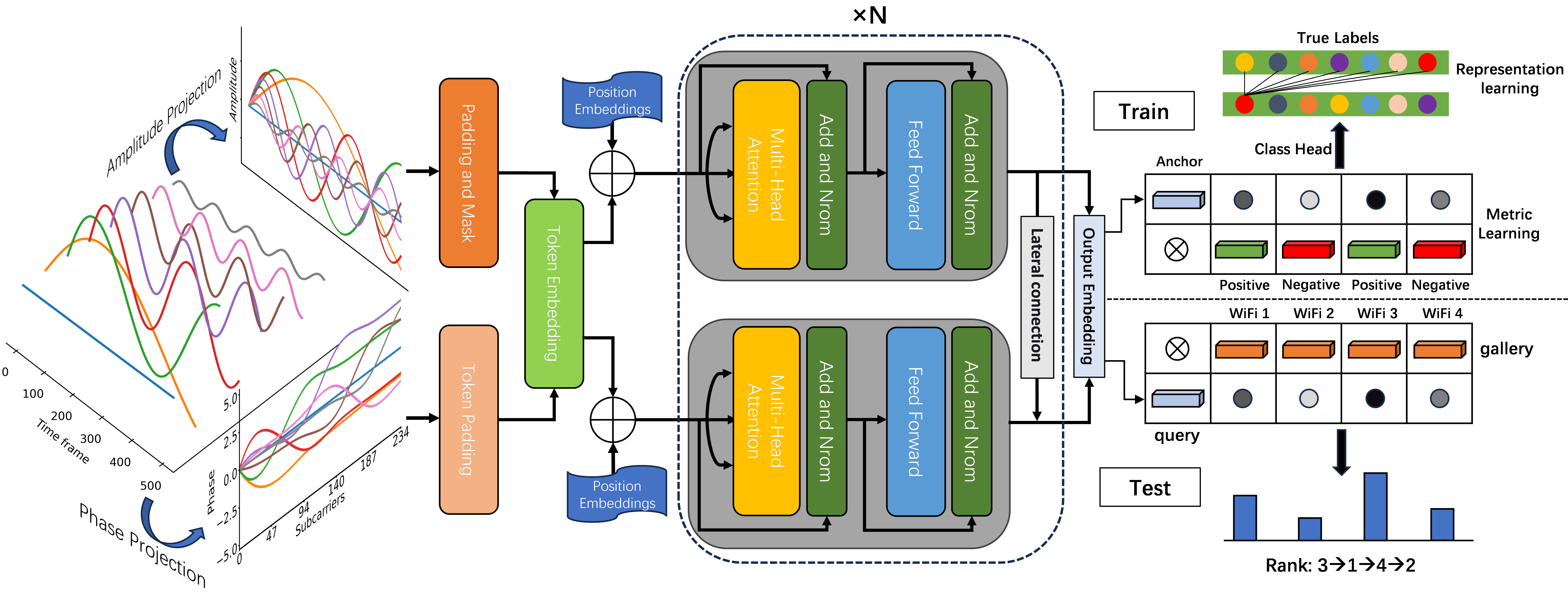}
    \caption{Architecture of the two-stream time-frequency person ReID network based on WiFi CSI signals.}
    \label{fig:wifi_arch}
    \vspace{-.35cm}
\end{figure*}

\begin{figure}[t!]
    \centering
    \includegraphics[width=\linewidth]{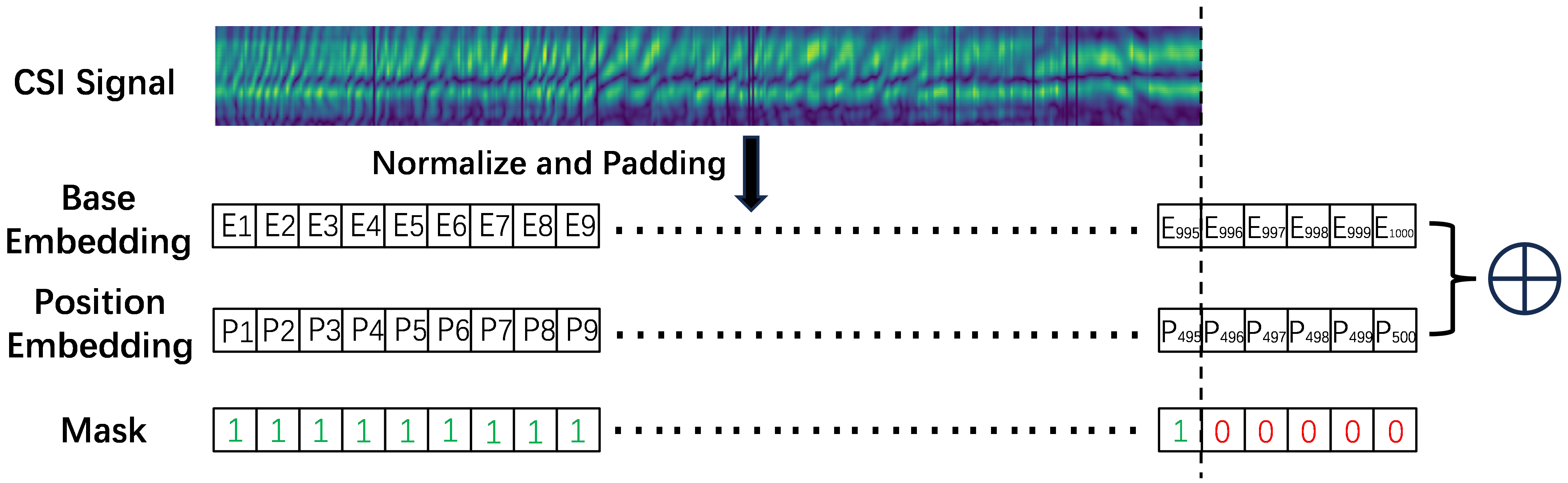}
    \caption{After normalizing and padding the length of the signal data, the base embeddings and position embeddings are merged to form a complete input.}
    \label{fig:csi_uniform}
    \vspace{-.35cm}
\end{figure}

\subsection{Two-Stream Network Architecture}
 
Given the variability in the length of pedestrian WiFi time-series data received in real-life scenarios, data length can differ significantly. If a pedestrian stays longer near the receiving router, the signal's duration will be longer. Truncating the data during training and testing processes without fully utilizing the information leads to wasted data potential. Inspired by the BERT\cite{2018BERT} in natural language processing models, we uniformly process the input WiFi CSI.

Analyzing signal amplitude from a time domain perspective, for tensors of varying length $(time frame, subcarriers)$, we set a uniform length in the time dimension. Sequences not meeting this length are zero-padded at the end to ensure all sequences reach the same length $(max time, subcarriers)$, maintaining input dimension consistency for batching data into the model for training and testing. The padding length is determined by a predefined maximum length, with truncation only occurring when a sequence exceeds this maximum length. To differentiate between real data and padding, a masking mechanism is introduced to indicate the positions of valid data. The mask values are typically set to 0 and 1, corresponding to padding and non-padding positions, respectively. When computing sequence self-attention in the feature extraction network, the mask is used to exclude attention to padding positions, ensuring the model focuses solely on real data, unaffected by padding.

From a frequency domain perspective, analyzing signal phase for tensors with fixed sequence length but variable token lengths $(subcarriers, time frame)$, we zero-pad the token dimension of the phase without needing to record a mask. This padding ensures token uniqueness, not affecting the model's ability to capture relationships between sequence signals' tokens. By integrating the mechanism, the model can efficiently process CSI sequence of any length in both time and frequency domains, significantly enhancing data utilization during training and inference.

To mark the temporal and frequency sequence of signals, after normalizing the signal data, we add learnable position embeddings on top of the base embedding in the sequence dimension. These are used to represent the position of each moment or frequency information within the WiFi CSI segment in the sequence. Subsequently, the time-domain amplitude and frequency-domain phase data are fed into a stacked two-stream transformer encoder\cite{vaswani2017attention} feature extraction network. This network uses a self-attention mechanism to address the long-distance dependency issues within the sequence. The architecture of the two-stream time-frequency person ReID network based on WiFi CSI signals is shown in Fig. \ref{fig:wifi_arch}, while the unified processing and position marking of variable-length WiFi signals are illustrated in Fig. \ref{fig:csi_uniform}.

\subsection{Time-Frequency Feature Fusion}


To facilitate the effective fusion of features from the amplitude and phase branches, we introduce a continuous lateral connection mechanism\cite{lin2017feature, 2019SlowFast}. This involves using lateral connections after the output of each pair of transformer encoders in the feature extraction network, merging features from the amplitude branch into the phase branch through element-wise addition. This structural design promotes the continuous flow of information between layers in the time and frequency domains and enables the model to perform feature fusion across multiple levels of high-level semantics and low-level details, thereby capturing more complex time-frequency features. The output from the lateral connections after the last layer of transformer encoder blocks serves as a comprehensive and complementary feature representation. This complete person feature representation captures not only the original strength and changes in the signal's propagation path but also considers the temporal dynamics and spectral details of the signal.

\subsection{Representation and Metric Learning}




We place the large margin cosine loss (LMCL)\cite{2018CosFace} behind the classification head of the model, aiming to enhance the representation capability of the model by maximizing intra-class similarity and inter-class dissimilarity. LMCL introduces cosine distance and a fixed margin to make feature vectors of the same class more compact, while increasing the distance between feature vectors of different classes. Based on the traditional softmax loss, LMCL normalizes the feature vectors and multiplies them by a scaling factor $s$ to increase the margin of the decision boundary. The specific formula is:
\begin{equation}
L_{\text{LMCL}} = -\frac{1}{N} \sum_{i=1}^{N} \log \frac{e^{s(\cos(\theta_{y_i}) - m)}}{e^{s(\cos(\theta_{y_i}) - m)} + \sum_{j \neq y_i} e^{s \cos(\theta_j)}}
\end{equation}


where $N$ is the batch size, $\theta_{y_i}$ is the angle between the feature vector of the correct class and the weight vector, $m$ is the introduced margin, and $s$ is the scaling factor.


We place SoftTriple\cite{2019SoftTriple}, a metric learning-based approach, at the feature output layer with the aim of more effectively learning the fine-grained structure of categories within the feature space. SoftTriple allows each category to have multiple centers to better handle intra-class diversity. This is achieved by introducing a soft assignment mechanism, which can attribute sample features to one of several centers within the same category, thereby optimizing intra-class similarity and inter-class dissimilarity. The specific formula is defined as follows:
{
\fontsize{8pt}{10pt}\selectfont
\begin{equation}
L_{\text{SoftTriple}} = \frac{1}{N} \sum_{i=1}^{N} \log \frac{1 + \sum_{j=1}^{C} \sum_{k=1}^{K} \exp(-\sigma(d(x_i, c_{jk}) - \delta))}{\exp(\lambda) + \sum_{j \neq y_i} \sum_{k=1}^{K} \exp(-\sigma d(x_i, c_{jk}))}
\end{equation}
}

where $N$ is the batch size, $C$ is the number of categories, $K$ is the number of centers per category, $d(x_i, c_{jk})$ is the distance between sample $x_i$ and center $c_{jk}$ of category $j$, $\sigma$ is the scaling parameter, and $\delta$ and $\lambda$ are parameters controlling the degree of separation between and within categories, respectively.

\section{Experiments}
\tit{Implementation Details} 


We utilize ViFi-Indoors\cite{githubGitHubMaomao279ViFiIndoors} as the train and test dataset for the method proposed in this letter. It is an unprocessed raw dataset containing WiFi information from 20 persons, collected using 234 subcarriers, and transmitted via a single antenna with four receiving antennas, forming a 1$\times$4 MIMO matrix. The frequency of the WiFi signal transmission is 100 times per second. We merge the dimensions of the receiving antennas and subcarriers, and randomly split them along the time dimension, resulting in 1339 segments of variable-length WiFi CSI signal data.

We adhere to the rules and paradigms established for ReID datasets, where data from the same individual can only be present in either the training set or the test set, but not both. We randomly selected WiFi data from 11 individuals for the training set, resulting in 738 signal segments, and WiFi data from 9 individuals for the test set, resulting in 601 signal segments. Within the test set, we sample some data as the query set, with the remaining data serving as the gallery set. For performance evaluation, we use the standard metrics of Rank-N, mAP, and mINP, which are commonly adopted in most ReID literature.

Our computational resource is a single RTX 4090 GPU, with a batch size of 16, the computation speed is 0.14 seconds per batch, allowing us to analyze 114 signals per second.

\tit{Data Augmentation} 
We propose several data augmentation techniques for WiFi signals, such as temporal distortion, which introduces nonlinear variations in the time dimension of the signal to simulate possible speed changes during signal transmission. By injecting random noise into the signals, we can simulate environmental interference. Additionally, randomly obscuring parts of the signal can mimic situations of signal loss or transmission errors. Introducing these variations during the training process can significantly enhance the model's generalization ability and robustness.

\tit{Model Comparison}


To demonstrate the effectiveness of the proposed model, we compare it with current mainstream network models. After training the models under the same experimental conditions, we evaluate multiple models using the same test set. As shown in TABLE~\ref{table:3}, our model achieves superior performance across all evaluation metrics compared to models such as ResNet.
\begin{table}[htbp]
\centering
\caption{Comparison of different network models.}
\label{table:3}
\begin{tabular}{ccccc}
\hline
Method & mAP$\uparrow$ & mINP$\uparrow$ & Rank-1$\uparrow$ & Rank-5$\uparrow$ \\
\hline
MobileNetv3\cite{9008835} & 85.45 & 75.20 & 91.30 & 94.07 \\
ResNet\cite{he2016deep} & 90.54 & 80.01 & 95.13 & 98.00 \\
ResNext\cite{xie2017aggregated} & 91.10 & 81.00 & 96.20 & 98.60 \\
ShuffleNet\cite{8578814} & 89.80 & 79.50 & 94.70 & 97.30 \\
LSTM\cite{shi2015convolutional} & 87.60 & 77.80 & 92.80 & 95.50 \\
Ours & \textbf{93.68} & \textbf{85.19} & \textbf{98.13} & \textbf{100.00} \\
\hline
\end{tabular}
\vspace{-.35cm}
\end{table}

\tit{Ablation Study}
To explore the impact of objective functions and data augmentation on the overall performance, this section conducts ablation experiments. We use a method that does not employ data augmentation, utilizing basic cross-entropy and triplet losses as the baseline, and then progressively incorporate the methods proposed in this letter.
\begin{table}[htbp]
\centering
\caption{Verifying the impact of data augmentation and advanced losses on each metric.}
\label{table:1}
\resizebox{\columnwidth}{!}{%
\begin{tabular}{ccccccc}
\hline
\multicolumn{3}{c}{Methods} & \multirow{2}{*}{mAP$\uparrow$} & \multirow{2}{*}{mINP$\uparrow$} & \multirow{2}{*}{Rank-1$\uparrow$} & \multirow{2}{*}{Rank-3$\uparrow$} \\
\cline{1-3}
                       Aug. & Sof./Tri. & LMCL/Cro. &                                   &                                   &                                   &        \\
\hline
                       - & Tri. & Cro. & 90.05 & 74.58 & 95.33 & 99.07  \\
                       \checkmark & Tri. & Cro. & 91.21 & 75.81 & \textbf{98.13} & \textbf{100.00} \\
                       \checkmark & \textbf{Sof.} & Cro. & 89.68  & 82.12 & 94.39 & 96.26 \\
                       \checkmark & Tri. & \textbf{LMCL} & 92.42 & 79.15 & \textbf{98.13} & 99.07 \\
                       - & \textbf{Sof.} & \textbf{LMCL} & 91.41 & 83.39 & 96.26 & 99.07  \\
                       \checkmark & \textbf{Sof.} & \textbf{LMCL} & \textbf{93.68} & \textbf{85.19} & \textbf{98.13} & \textbf{100.00}  \\
\hline
\end{tabular}%
}
\vspace{-.35cm}
\end{table}

As seen in TABLE~\ref{table:1}, the inclusion of data augmentation and advanced losses led to improvements of 10.62\% in mINP and 2.08\% in mAP, metrics that represent model robustness, indicating that by using corresponding objective functions at different output layers of the model, not only can the model learn distinctive feature representations, but it can also adapt to intra-class variability. This significantly improves the model's feature clustering ability, thereby achieving better performance in the ReID task. Data augmentation, by introducing disturbances to the original signal, reduces the frequency at which the model learns from the original signal, leading to no improvement in the Rank-N metric, but significantly enhances the model's robustness.

\begin{figure}[htbp]
    \centering
    \includegraphics[width=\linewidth]{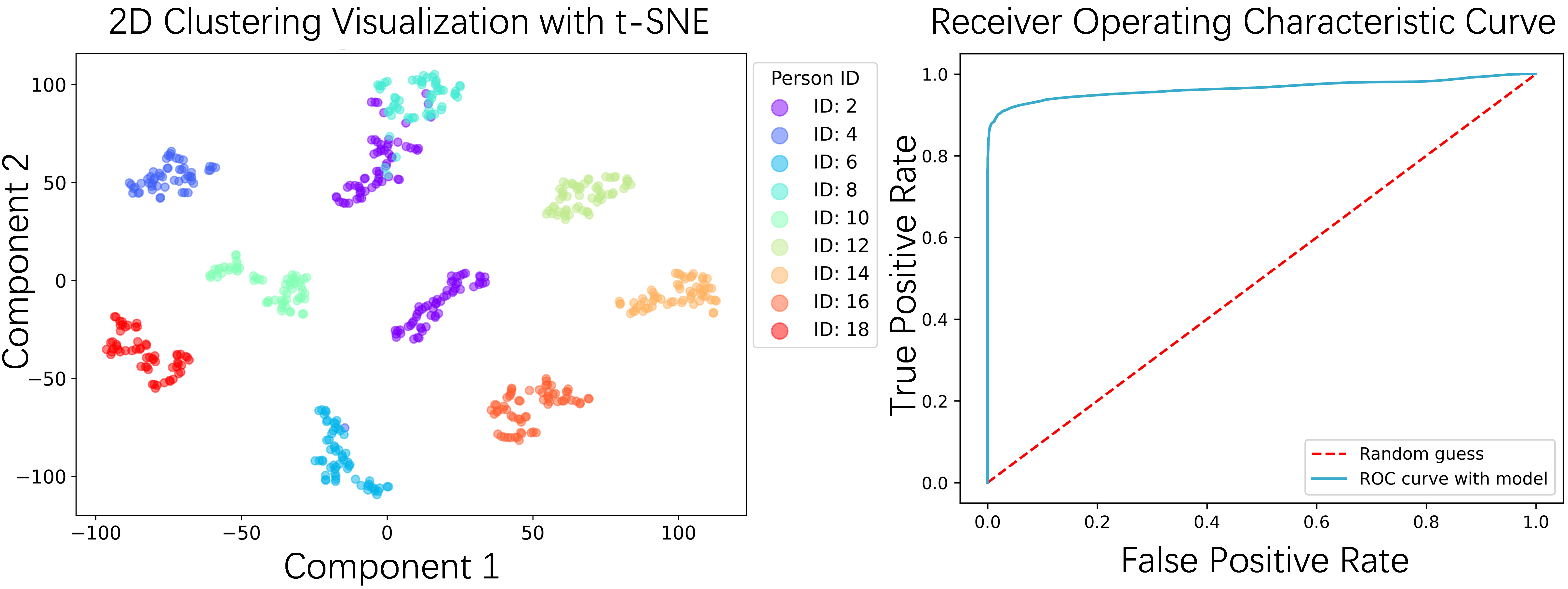}
    \caption{Visualizations of pedestrian feature clustering in two-dimensional space and ROC curve.}
    \label{fig:clustering}
\end{figure}

The effect of reducing the pedestrian features output by the model to a two-dimensional space using the t-SNE clustering algorithm\cite{van2008visualizing}, along with the ROC curve, is shown in Fig. \ref{fig:clustering}. It can be observed that the method is capable of significantly clustering the same pedestrian features in the feature space, and the model can decisively distinguish between positive and negative classes, demonstrating excellent recall and precision rates.

\vspace{-.35cm}

We also conduct ablation studies on different feature fusion methods, comparing the continuous lateral connection used in our model with mainstream early and late fusion methods. TABLE~\ref{table:2} indicates that our method is more capable of integrating information from both the time and frequency domains across multiple levels, thereby achieving a comprehensive representation of pedestrian features.
\begin{table}[htbp]
\centering
\caption{Comparison of different feature fusion methods.}
\label{table:2}
\begin{tabular}{ccccc}
\hline
Fusion & mAP$\uparrow$ & mINP$\uparrow$ & Rank-1$\uparrow$ & Rank-5$\uparrow$ \\
\hline
Early Fusion & 93.45 & 82.70 & 97.20 & 99.07 \\
Late Fusion & 92.54 & 84.01 & \textbf{98.13} & \textbf{100.00} \\
CLS (Ours) & \textbf{93.68} & \textbf{85.19} & \textbf{98.13} & \textbf{100.00} \\
\hline
\end{tabular}
\vspace{-.35cm}
\end{table}

\section{Conclusion}


In this letter, we propose a method for person re-identification using WiFi signals. By employing a dual-stream network to analyze the temporal amplitude and frequency domain phase of WiFi CSI signals, and utilizing continuous lateral connections for multi-level time-frequency feature fusion, this non-visual information-based approach demonstrates the potential of leveraging existing wireless network infrastructure for intelligent surveillance. This method provides a new research direction and application possibilities for person re-identification and security.

\bibliographystyle{IEEEtran}
\bibliography{bibliography}
\end{document}